\documentclass[twoside]{article}
\usepackage{fleqn,espcrc2}

\usepackage{epsfig}
\title{Computers for Lattice QCD}

\author{Norman H. Christ\address{Department of Physics, 
        Columbia University, New York, NY 10027 USA}
        \thanks{This work was supported in part the 
        U.~S.~Department of Energy, under grant \# DE-FG02-92RE40699.}}
       
\begin{document}

\begin{abstract}
The architecture and capabilities of the computers currently in use for 
large-scale lattice QCD calculations are described and compared.  Based 
on this present experience, possible future directions are discussed.
\vspace{1PC}
\end{abstract}
\maketitle

\section{INTRODUCTION}

The overriding objective of present work in lattice QCD is to achieve an 
accurate representation of continuum QCD and to use this ability to study 
physically important properties of the underlying theory.  This research
demands a combination of efficient numerical algorithms, physical quantities
appropriately represented for numerical study and fast computers to carry
out the needed calculations.  While the development of algorithms and
inventions of methods to tackle new problems are natural tasks for 
theoretical or computational physics, it is most often the large machines
that produce the actual numerical results of importance to physics.  
Thus, the development, characteristics and availability of large scale
computer resources are necessarily central topics in lattice QCD.

Fortunately there is much progress to report since this topic was last
addressed in a plenary talk at a lattice meeting\cite{SEXTON}.  In
particular, since that review the 300 Flops PC-PACES in Tsukuba and 
120 and 180 Gflops QCDSP machines at Columbia and the RIKEN Brookhaven 
Research Center have come into operation.  In addition construction
is beginning on the next APE machine, APEmille.  These large machines
offer critical opportunities for progress on the most demanding 
calculations at the frontier of lattice QCD, especially exploration
beyond the quenched approximation in which the full effects of dynamical fermions are incorporated.

While a survey of present computer resources (see below) shows a marked
decline in the importance of large-scale commercial supercomputers 
(only the Hitachi machine in Tsukuba belongs to this category), there
is a promising appearance of a new class of machines: workstation farms.
While also commercial machines, workstation farms are often assembled
by the group wishing to use them and are built from cost effective
PC's or workstations connected with a commercial network.  While present
workstation farms cannot deliver the high performance offered by the 
large projects mentioned above, they are easily assembled and can 
quickly exploit advances in PC technology.

In Table~\ref{tab:survey} we follow Sexton's example and attempt 
to give a picture of the computer resources presently available for
lattice QCD research.
\begin{table*}[htb]
\caption{A summary of computer resource presently available for
lattice QCD.  These are rough estimates which attempt to quantify 
available computer resources in average, sustained Gigaflops.}
\label{tab:survey}
\begin{center}
\begin{tabular}{ll|lr|lr}
\hline
\multicolumn{2}{c|}{}    &\multicolumn{2}{c|}{Lattice '95}
                                  &\multicolumn{2}{c}{Lattice '99} \\
Country	& Group	& Machine & Power
							& Machine & Power  \\
\hline
\underline{Germany:}
		& various	& 3QH2    & 21      & QH4+QH2   & 21       \\
\underline{Italy:}
		&RomaI	& 4QH4    & 55      & 4QH1+2QH2 & 28       \\
		&RomaII	&  QH4    & 14      & 1QH1      & 3        \\
		&Pisa		& QH4+2QH1& 21      & 2QH1      & 7        \\
		&Other	& 2QH1    & 7       & NQH1      & 16       \\
\underline{Japan:}
		&Tsukuba	& various & 10      & PC-PACES	  & 300	 \\ 
            &JLQCD	& VPP500(80)& 50    & VPP500(80)& 50       \\ 
            &RIKEN	& VPP500(30)& 10    & VPP800E   & 30       \\ 
\underline{UK:}
		& UKQCD	& various   & 2.5   & T3E/152   & 41       \\ 
\underline{US:}
		& Columbia	& 256-node  &7      & QCDSP     & 120      \\ 
            &RIKEN/BNL	& ---       &       & QCDSP     & 180      \\ 
            &Fermilab   & ACPMAPS   & 5     & ACPMAPS   & 5        \\ 
            &IBM 		& GF11      & 5	  & ---       &          \\ 
         	&LANL     	& various   & 5	  &Origin2000 & 3.6      \\ 
            &MILC       & various   & 5	  & various   & 6		 \\ 
\hline \\
\end{tabular}
\end{center}
\vskip -0.2in
\end{table*}

\section{THREE LARGE PROJECTS}

The PC-PACES, QCDSP and APEmille machines represent somewhat varied 
examples of how large-scale computer resources can be provided for
lattice QCD studies.

\subsection{PC-PACES}

The first of this generation of machine to be completed was 
PC-PACES\cite{PC-PACES} which has been working since the middle 
of 1996.  It contains 2048 independent processors connected by a 
3-dimensional hyper-crossbar switch joining processors and
I/O nodes into a $17\times 16 \times 8$ grid.  All nodes
with two out of three identical Cartesian coordinates are connected
by their own 300 MB/sec crossbar switch.

The processor is an enhanced version of a standard Hewlett 
Packard reduced instruction set computer and carries out 
64-bit floating point arithmetic.  The processor enhancement involves
the addition of extra floating point registers allowing the efficient 
transfer of long vectors into the processor. The peak speed of 
each processor is 300 Mflops so the peak speed of the 2048-node 
machine is 0.6 Teraflops.

The resulting architecture achieves remarkable efficiencies for QCD code 
(above 50\%\cite{PC-PACES-perform}) and appears to be useful to other
applications as well.  Perhaps equally impressive is the fact that
FORTRAN, $C$ and $C$++ compliers are available which support the 
high performance features of the machine.  While much of the lattice
QCD code is written in these high level languages, careful assembly
language programming is used for the critical inner loops.  

The machine consumes 275 KWatts and was built within a \$22M budget
giving a cost performance figure of \$73/Mflops.  The machine can be 
partitioned into a number of disjoint units, each with its own queue.
Typically the queues are reconfigured about once per month.  During 
their large quenched hadron mass calculation the machine was 
configured as a single partition about one third of the time.

\subsection{QCDSP}

\begin{figure}[t]
\vskip -0.1in
\epsfxsize=2.9in
\epsfbox{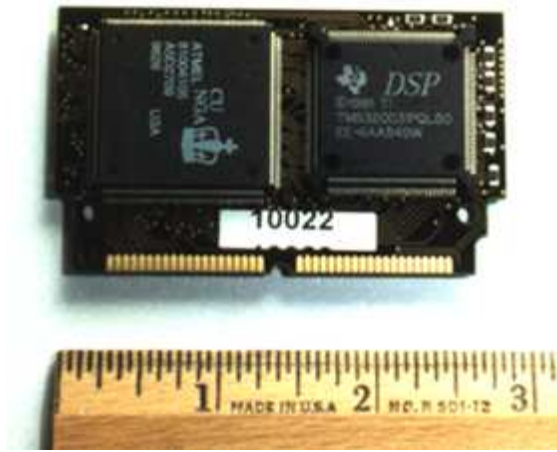}
\vskip -.2in
\caption{Picture of a single QCDSP node.  The gate array chip
is on the left and the DSP on the right.  The five memory
chips are on the back side.}
\label{fig:node}
\vskip -0.3in
\end{figure}

The next machines of this present generation to be finished are the 
QCDSP machines at Columbia and the RIKEN Brookhaven Research 
Center\cite{QCDSP}.  These machines are based on a very 
simple node made up of a commercial, 50 Mflops, 32-bit digital 
signal processor chip (DSP), a custom gate-array (NGA) and 2 MB 
of memory.  These are mounted on a single 4.5$\times$7.5cm SIMM 
card as shown in the picture in Figure \ref{fig:node}.  The NGA 
enhances DSP performance by providing a 32-word buffer 
between the DSP and memory.  This buffer acts in different ways 
depending on which of nine images of memory the DSP addresses, 
varying from normal memory access (with a minimum 
of 2 wait-states) to a fetch ahead mode which permits 25 Mword/sec 
data motion from memory and 0 wait-state random access by the DSP 
to a portion of the 32-word buffer.

The NGA also provides inter-processor communication, supporting 
simultaneous $\approx$40Mb/sec serial communication between the node 
and each of its eight nearest neighbors in four dimensions.  This 
net 40 MB/sec bandwidth per node is sufficient for efficient 
execution of lattice QCD code even when there are as few as $2^4$ 
sites per node.  In addition, the NGA implements pass-through or
``worm-hole'' operations that allow efficient global floating
point sums and broadcasts.

Most of the code for these QCDSP machines is written in $C$++ with 
the critical low level routines written in DSP assembly language.
The fastest code for Wilson or domain wall fermions achieves 30\% 
of peak performance or 15 Mflops on a single node.  Thus the 8,192
node machine at Columbia and the 12,288 node machine at the RBRC
at Brookhaven can achieve 0.12 and 0.18 Teraflops respectively.
The RBRC machine was the final machine constructed and, including
Brookhaven assembly labor, cost \$1.85M giving a cost performance 
figure of \$10/Mflops and received the 1998 Gordon Bell Prize 
for cost performance.  It consumes approximately 60 KWatts.
By a rearrangement of cables, a QCDSP machine can be configured
as a single large machine run from a single host to variety of
smaller machines run by many hosts, with some hosts running an
array of machines.

\subsection{APEmille}

APEmille is the next in a series of QCD machines built by the Italian 
and now DESY groups\cite{APE}.  This machine is constructed of 
custom floating point processors, with a peak, single-precision 
speed of 538 Mflops.  Each processor has 32 Mbytes of memory 
and is connected to its six nearest neighbors in a 3-dimensional 
mesh.  The total off-node bandwidth provided to a single processor
is 132 MB/sec.  The processors are organized into clusters of 
eight with each such cluster fed instructions SIMD-style by a single 
controller.  Four of these eight-node clusters are then controlled
by a built-in Linux ``workstation'' with I/O capability.  One large 
machine can be subdivided in software into a number of independent 
SIMD units.

The machine will be programmed in two languages $C$++ and TAO, with
the later language providing backward compatability with APE100 TAO
code.  A 2048-node machine is expected to consume 20 KWatts.  Such a machine
would have a peak speed of 1 Teraflops and is expected to sustain at
least 50\%, or 0.5 Teraflops.  At a cost of \$2.5M this will yield a
machine with a cost performance of \$5/Mflops.  There is a 32-processor 
unit working now.  Two 250 Gflops and two 128 Gflops units are planned 
for INFN and at least one 250 Gflops unit for DESY, all by next summer.

\section{SMALLER MACHINES}

In addition to the three large projects described above, there are a 
number of smaller machines that provide important resources for
lattice QCD calculations.  The UKQCD collaboration has the use of a
152-node T3E.  With 64 Mbyte, 0.9 Gflops nodes, this machine has 
a peak performance of 137 Gflops and sustains 41 Gflops on QCD code.
Usually there is a large, 128-node partition devoted to
production running.  There is a 128-node Fujitsu VPP700E with 2 Gbyte,
2.4 Gflops nodes at RIKEN.  This machine sustains 150 Gflops and is
perhaps 20\% available for lattice calculations.  An 80-node, 80 
Gflops VPP500 is installed at KEK with an upgrade to a 550 Gflops 
sustained machine expected early next year.  In the U.S., Fermilab 
continues to use its ACP-MAPS machine which sustains 5 Gflops.  The
LANL group has access to a 64-node 32 Gflops Origin2000 which sustains
3.6 Gflops for QCD.  Finally the MILC group has access to a variety of
machines include an Origin2000, a T3E and an SP-2, yielding a total
sustained performance of 5.8 Gflops.

\section{WORKSTATION FARMS}

An important recent development is the appearance of clusters of 
workstations, networked together to tackle a single problem.  
This approach to parallel computing has the strong advantage that 
it can be pursued with commodity components that are easily 
assembled and uses standard, platform-independent software, 
{\it e.g.} Linux and MPI.  We consider four examples:

\underline{Indiana University Physics Cluster.}  Each node in 
this 32-node cluster is a 
350MHz Pentium II with 64 MB of memory, joined through a 
switch using 100MB/sec Fast Ethernet.  Running a straight 
port of MPI MILC code, S. Gottlieb reports the following 
benchmarks.  A staggered inverter applied to a $4^4$ lattice on a 
single node, yields 118 Mflops while for a $14^4$ lattice the 
performance falls to 70 Mflops (presumably due to less efficient 
cache usage).  Identical code run on the entire 32-node machine
achieves 0.288 Gflops on a $8^3\times 16$ lattice ($4^4$ sites 
per node).  This number increases to 1.4 Gflops for a $28^3\times 
56$ lattice ($14^4$ sites per node).  Thus, the effects of the 
somewhat slow network are to decrease the single node 118 Mflops
to 9 Mflops for the smaller lattice while for the large lattice
of $14^4$ sites per node the decrease is less severe: 70
to 44 Mflops.  While the Fast Ethernet is too slow to allow maximum
efficiency, it is very inexpensive, allowing a total machine cost
of \$25K in Fall of '98 and impressive cost performance numbers of 
\$87 and \$18/Mflops for the two cases.

\underline{Roadrunner Cluster.} This is 64-node machine at the 
University of New Mexico with each node made up of two Pentium II, 
450~MHz, 128~MB processors for a cost of approximately \$400K.
S. Gottlieb's benchmarks (using only one of the two processors) 
show single-node performance of 127 and 71 Mflops for the $4^4$ 
and $14^4$ cases. On a 32-node partition, he finds 1.25 
and 2.0 Gflops for the $8^3\times 16$ and $28^3 \times 56$ 
examples.  This cluster has a much faster network using 
Myrinet (1.28 Gb/sec) and a much better 39 Mflops
performance per node on the small lattice with the greatest 
communication demands.  The higher performance is balanced by
the higher cost of this commercial machine.  Arbitrarily 
discounting the \$400K by 1/3 since only one of two 
processors was used, we obtain \$106/Mflops and \$67/Mflops 
for these two different sized lattices.

\underline{ALICE.}  This is an 128-node, Alpha-based machine 
with a Myrinet multistage crossbar network being assembled in 
Wuppertal.  Each node is a 466 MHz 21264 Alpha processor with 
single-node performance for QCD code of 175 Mflops
(double precision).  The full system is expected achieve
20 Gflops (double) and $>$30 Gflops (single precision).  Based
on a list price of about \$0.9M, this machine provides \$45/Mflops,
a number sure to be lower when the actual price is determined.

\underline{NCSA NT Myrinet Cluster.}  This cluster of 96, dual 
300MHz Pentium II nodes has been used by D.~Toussaint 
and K.~Orginos for MILC benchmarks and by P.~Mackenzie and 
J.~Simone for a Canopy test, each using a $12^3\times 24$ 
lattice.  While the performance per node decreases from 22.5 
to 12.8 for the Canopy test as the number of nodes increases
from 1 to 12, the MILC code runs at essentially 50 Mflops/node
as the number of nodes varies between 1 and 64.  Thus, a 64-node
machine sustains 3.2 Gflops for the MILC code.  It is impressive
that the Canopy code could be ported without great difficulty
and its performance is expected to increase as extensions to MPI are
added which provide features needed for the efficient execution 
of Canopy.

\section{THE FUTURE}

\begin{figure}[t]
\vskip -0.6in
\epsfxsize=3.1in
\centering
\hskip -.2in \epsfbox{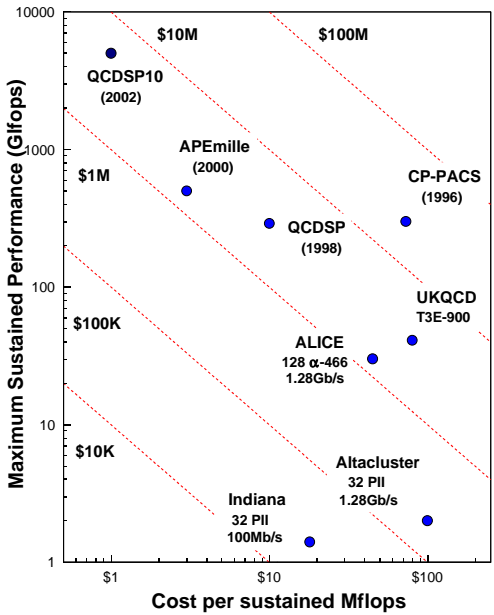}
\vskip -.6in
\caption{Total sustained performance plotted against price
performance for a number of systems used for lattice QCD.  The diagonal
lines are lines of constant system cost.}
\label{fig:perf}
\vskip -0.3in
\end{figure}

In light of the above discussion it is interesting to discuss
possible future directions for lattice QCD computing.  
Figure~\ref{fig:perf} shows the computing power and price performance 
of the various systems just discussed.
That summary suggests that workstation farms will soon offer
unprecedented, cost-effective, 10-50 Gflops computing.  By allowing
rapid upgrades to current technology and using standard
software, this approach provides an easy migration path to 
increasingly cost effective hardware.  If a relatively fast network is
provided, the general interconnectivity supports a variety of
communication patterns which are easily accessible from portable 
software.  Workstation farms should provide increasingly accessible
and powerful resources for lattice QCD calculations.

However, the largest size of a farm that can be efficiently 
devoted to evolving a single Markov stream is limited 
by communication technology.  Commodity Ethernet is 
too slow and the more effective 1.28 Gb/sec Myrinet 
product, at \$1.6K/node, does not benefit from a mass market.  
In addition, the intrinsic latency of a general purpose network 
product, limits its applicability for lattice QCD.  For a 
500 Mflops processor, 10 $\mu$ sec is required for the 
application of a single staggered $D$ operator to the even sites 
of a $2^4$ local lattice.  However, the smallest latency achieved by a 
Myrinet board is 5 $\mu$s and sixteen independent transfers are 
needed for such a $D$ application implying a minimum 8$\times$ 
performance decrease for such a demanding problem. 

Thus, it seems likely that high end computing, needed for 
example to evolve necessarily small lattices with dynamical quarks, 
will be done on purpose-built machines of the PC-PACES, APEmille, 
QCDSP variety.  If fact, all three groups are already planning 
their next machine.  While no details are available from Tsukuba, 
the APE machine, APEnext, aims for 3-6 Tflops peak speed with 
64-bit precision, 1 Tbyte of memory and 1 Gb/sec disk I/O.  The next
Columbia machine, QCDSP10, is planned as an evolution of the QCDSP
architecture.  The new processor may be a 0.67~Gflops, C67 Texas 
Instruments DSP with 128 KB of on-chip memory.  The chip also 
supports 64-bit IEEE floating point at 0.17 Gflops.  We plan to begin 
development this Fall and hope to have a 16K node, 10 Tflops peak, 
5 Tflops sustained machine in operation within three years.  In 
addition, we would like to construct a larger, community machine 
supporting a more competitive scale of high end computing for the 
entire U.S. lattice QCD activity.

\section{ACKNOWLEDGMENTS}

The author is indebted to many people for providing much of the 
information presented here, in particular, S.~Gottlieb, R.~Gupta, 
B.~Joo, K.~Kanaya, R.~Kenway, T.~Lippert, P.~Mackenzie, 
R.~Petronzio, S.~Ohta, M.~Okawa, J.~Simone, D.~Toussaint, R.~Tripiccione, 
and T.~Yoshie.

\end{document}